# Global Trends and Predictors of Face Mask Usage During the COVID-19 Pandemic

Elena Badillo-Goicoechea, MS[1*], Ting-Hsuan Chang, BS[2*], Esther Kim, PhD[3], Sarah LaRocca, PhD[3], Katherine Morris, PhD[3], Xiaoyi Deng, MS[4], Samantha Chiu, MA[4], Adrianne Bradford, MS[4], Andres Garcia, BS[4], Christoph Kern, PhD[5], Curtiss Cobb, PhD[3], Frauke Kreuter, PhD[4,5,6], Elizabeth A. Stuart, PhD[1,2]

[1] Department of Mental Health, Johns Hopkins Bloomberg School of Public Health, Maryland, USA
[2] Department of Biostatistics, Johns Hopkins Bloomberg School of Public Health, Maryland, USA
[3] Facebook Research, California, USA
[4] Joint Program in Survey Methodology, University of Maryland College Park, Maryland, USA
[5] Department of Statistics, Ludwig Maximilians University, Munich, Germany
[6] Institute for Employment Research, Nuremberg, Germany

* Contributed equally as first authors.

**Running title**: Global trends and predictors of face mask usage

**Word count**: Abstract (300), main manuscript excluding figures, tables, and acknowledgements (3509)

**Corresponding author**:

Elizabeth A. Stuart, PhD

**Address**:
Johns Hopkins Bloomberg School of Public Health
615 N. Wolfe St. W1033
Baltimore, MD 21205

**Email**:
estuart@jhu.edu

**Phone**: 410-955-9088

## ABSTRACT

**Background**: Guidelines and recommendations from public health authorities related to face masks have been essential in containing the COVID-19 pandemic. We assessed the prevalence and correlates of mask usage during the pandemic.

**Methods:** We examined a total of 13,723,810 responses to a daily cross-sectional representative online survey in 38 countries who completed from April 23, 2020 to October 31, 2020 and reported having been in public at least once during the last seven days. The outcome was individual face mask usage in public settings, and the predictors were country fixed effects, country-level mask policy stringency, calendar time, individual sociodemographic factors, and health prevention behaviors. Associations were modelled using survey-weighted multivariable logistic regression.

**Findings**: Mask-wearing varied over time and across the 38 countries. While some countries consistently showed high prevalence throughout, in other countries mask usage increased gradually, and a few other countries remained at low prevalence. Controlling for time and country fixed effects, sociodemographic factors (older age, female gender, education, urbanicity) and stricter mask-related policies were significantly associated with higher mask usage in public settings, while social behaviors considered risky in the context of the pandemic (going out to large events, restaurants, shopping centers, and socializing outside of the household) were associated with lower mask use.

**Interpretation:** The decision to wear a face mask in public settings is significantly associated with sociodemographic factors, risky social behaviors, and mask policies. This has important implications for health prevention policies and messaging, including the potential need for more targeted policy and messaging design.

**Funding**: This work was supported in part by the National Science Foundation NSF-RAPID award "Evaluating the Impact of COVID-19 on Labor Market, Social, and Mental Health Outcomes" (Bennett, de Bruin, Kreuter, Stuart, Thrul), and in part by Facebook for survey data collection and support time for data analysis.

## INTRODUCTION

In an effort to control and prevent the spread of the novel coronavirus disease 2019 (COVID-19), health organizations have recommended the use of a face covering or mask in public settings, and an increasing number of studies suggest that face masks may be effective in reducing the transmission of COVID-19.[1–6] Yet, despite growing evidence of the effectiveness of using face masks,[7] there is still a lack of data and formal studies examining mask-wearing behavior on a global scale. First, it is still unclear how mask-wearing behavior has changed over time across the globe. Second, it is unknown whether individual- or country-level factors are associated with mask-wearing. These questions are critical to better understand and target risky behavior patterns across individuals and places, clarify ongoing public health messaging around face mask usage, and, more generally, help better design prevention campaigns in future public health emergencies.

Very few studies have examined the trends and predictors of face mask usage during the COVID-19 pandemic. Increased usage of face masks has been reported in a few countries since Spring 2020,[6,8,9] and some studies have examined sociodemographic factors and individual beliefs and attitudes as predictors of mask-wearing, but mostly in the context of past outbreaks such, as SARS-Cov-1 and H1N1.[10–14] Additional limitations of previous work are that many used small non-random samples (e.g., ~300-5000 self-selected participants); most had a limited time frame (e.g., one or two months) and/or narrow geographical coverage (e.g., one country); and many studies could not simultaneously examine individual- and country-level factors that may explain differences in mask-wearing behavior.

The main objective of this study was, therefore, to examine the evolution of mask usage across different countries over time during the COVID-19 pandemic and assess whether individual and country-level factors were associated with the decision to wear a mask. We leveraged a novel dataset from the COVID-19 Symptom Survey,[15] which since April 2020 has daily tracked mask usage, sociodemographic characteristics, and other health prevention behaviors. We used data from respondents who were randomly selected to take a survey between April 23, 2020 to October 31, 2020 and who reported having been in public at least once during the last seven days, which included approximately 13 million adults in 38 countries.

To the best of our knowledge, the COVID-19 Symptom Survey is currently the largest data collection effort systematically monitoring mask usage and other social responses to the ongoing COVID-19 pandemic at a global scale with representativeness at a country-level. No other study has used its data to formally assess trends and predictors of mask usage worldwide.

## DATA AND METHODS

### The COVID-19 Symptom Survey

The COVID-19 Symptom Survey is an ongoing repeated daily cross-sectional survey conducted by the University of Maryland and Carnegie Mellon University in partnership with Facebook, Inc., and asks various questions related to symptoms, testing, preventive behaviors, mental health, and more. The international version of the survey was launched on April 23, 2020 in

>200 countries and territories (**Supplementary Table 1**) and the US version of the survey was launched on April 6, 2020. The survey instrument, sampling design, and weighting methodology are described in more detail below. The Symptom Survey was reviewed and approved by the Institutional Review Boards of both the University of Maryland and Carnegie Mellon University.

The survey instrument was developed by public health and survey experts,[15] and included the following sections: COVID-19 related symptoms, testing, contact history, preventive behavior (e.g., face mask usage, hand washing, social distancing, etc.), mental health, economic security, and basic demographics. The questionnaire is publicly available online,[16,17] and is translated into 56 locales (listed in **Supplementary Table 1**).

The sampling methodology for the Symptom Survey has been described previously.[18] Briefly, the sampling frame is composed of daily active Facebook users who are >=18 years, living within 200+ countries or territories, and using one of the supported languages. This coverage ensures that >95% of Facebook users are eligible. Every day, the Facebook app invites a stratified random sample to take the survey with an invitation at the top of their News Feed, with the sampling strata defined as the administrative boundaries within countries or territories.[19] Those who view the invitation and are interested in taking the survey are redirected to an off-Facebook survey administered by the academic institutions. Facebook does not share or receive data from the academic partners other than a list of random identification numbers of those who completed the survey to calculate and share survey weights.

The details of the weighting methodology have been described previously.[18] Briefly, Facebook employs a two-stage weighting process to minimize bias related to non-response and coverage. In the first step, inverse propensity score weighting is used to adjust for non-response bias by making the sample more representative of the sampling frame of Facebook users. As stated above, Facebook only receives a list of identification numbers that indicate who completed the survey; therefore, the covariates used in this step are obtained from internal Facebook data, which consist of self-reported age, gender, geographical variables, and other attributes that have been found internally to correlate well with survey response.[20] At the second stage post-stratification or raking is used to equate the distribution of age and gender among the Facebook population to benchmarks from the United Nations Population Division 2019 World Population Projections, and first administrative level region benchmarks from publicly available population density maps.[21]

**Study Population**

This analysis included adult participants who responded to the international COVID-19 Symptom Survey from April 23rd, 2020 until October 31st, 2020. We did not include responses from the US Symptom Survey in this analysis, as the question on face mask usage was not incorporated into the US questionnaire until September 2020. Since some of the 200+ countries and territories have relatively small sample sizes, with high variability in responses, we focused on 38 countries based on the following criteria: countries that are considered either members, candidates, or key partners of the Organisation for Economic Co-operation and Development (OECD) convention,[22] or countries with a sample size >600,000 during our study period (**Table 1**). Over the course of field collection in the selected 38 countries, 741,496,298 Facebook users saw the survey invitation; 36,525,312 opened the survey invitation; and 18,730,575 responded to the survey. Of those, 1,020,188 reported being in public in the past 7 days. Missingness on the predictors ranged from 2-13% per variable, which overall resulted in 27% of the survey respondents being excluded, leading to a final analysis sample of 13,723,810.

**Outcome Variable Definition**

Our outcome was face mask usage, based on the survey question: "In the last 7 days, how often did you wear a mask when in public?" The response options were "All of the time", "Most of the time", "Some of the time", "A little of the time", "None of the time", or "I have not been in public during the past 7 days". We defined face mask usage as a binary variable: 1 if the respondent reported wearing a mask all or most of the time, and 0 otherwise.

**Predictor Measurement**

We included several individual and country-level factors that could be associated with face mask usage based on a priori hypotheses and existing literature. Individual-level predictors included age, gender, standardized years of education, urbanicity (defined as living in a city versus town, village, or rural area), and the following reported social behaviors from the last 24 hours: working outside the household, going to a market/grocery store/pharmacy, going to a restaurant/cafe/shopping center, spending time with someone outside their household, and attending a public event with more than 10 people. We also included whether the respondent reported ever being tested for COVID-19, and two variables capturing individual economic aspects: worried about household finances and worked in the last seven days. The three variables on years of education, financial worry, and employment status in the last seven days were added to the survey on June 27, 2020; therefore, data on these items were not available earlier than this date.

Country-level predictors were country fixed effects, the (time-varying) presence of official policies related to face masks, and the (time-varying) incidence of COVID-19 disease. The country-level mask usage policies were obtained from the University of Oxford Our World in Data's COVID-19 dataset, which contains daily country-level policies on the use of face coverings outside-of-the-home. The policies are graded from 0-5 and reflect the strength of the policy (i.e., no policy, recommended, required in some specified places, required in all shared/public spaces, required at all times) for each country.[23] We generated standardized weekly averages of this mask-wearing policy stringency index for each country, and included the index as a continuous variable in the model. Country-day-level COVID-19 cases were obtained from the Johns Hopkins University Center for Systems Science and Engineering's repository,[24] which we used as a standardized seven-day lagged average to measure the association between the rate of COVID-19 cases during the last seven days and the individual's decision to wear a mask.

**Statistical Analysis**

In addition to examining descriptive statistics, a survey-weighted multivariable logistic regression model was used to formally assess whether individual and country-level factors were associated with mask-wearing. All statistical analyses were performed in R (version 4.0.3), using the R survey package (version 4.0) to account for the sampling design. We estimated two separate models to accommodate the fact that the three questions capturing socioeconomic factors (financial worry, years of education, and employment status in the last 7 days) were added later in field collection. The primary model included the entire sample from April 23, 2020, through October 31, 2020, with all predictors described above except for the three not available before June 27, 2020. A secondary model was fit with a narrower time period spanning from June 27, 2020 until October 31, 2020, to include the additional three socioeconomic factors. We included month as a categorical variable in all models.

**Role of the Funding Source**

Facebook provided a gift that partially supported the collection of survey data and time for data analysis. Facebook provided feedback on the manuscript, but all analyses were conducted independently by the researchers at Johns Hopkins University.

## RESULTS

### Sample Characteristics

**Table 1** documents the countries represented in the analysis sample and their corresponding sample sizes. The countries with the largest sample sizes were Brazil, Mexico, Japan, Italy, and India. Table 2 provides characteristics of the respondents used in analyses. Throughout the data collection period, most indicated that they had not gone to their workplace outside of home (64%), restaurant/cafe/shopping center (75%), or attended a public event with more than 10 people (89%) in the last 24 hours but had gone to a market, grocery store, or pharmacy in the last 24 hours (66%). The majority indicated that they wore a face mask most of the time or all of the time when in public (84%).

### Evolution of Mask Usage Over Time

Trends over time across the 38 countries (**Figure 1)** suggested considerable heterogeneity in self-reported mask-wearing in public across countries. Some countries had consistently high mask usage (>75%) from April until October (ex: Chile, Italy, Japan, Argentina, Colombia, Turkey, Romania, etc.) (**Figure 1A**). In some other countries, mask usage was relatively low in April, but eventually increased and remained at higher levels (ex: Brazil, Portugal, South Africa, Germany, France, Belgium, Greece, Canada, etc.) (**Figure 1B**). Mask usage was consistently low (<25%) in certain countries (ex: Denmark, Sweden, and Norway) (**Figure 1C**), and was more irregular in others (ex: Austria, Czech Republic, Slovenia, etc.) (**Figure 1D**).

### Predictors of Mask Usage

Results from the logistic regression model confirmed the observed cross-country mask-wearing trends over time, with individuals from the vast majority of countries — particularly of Northern Europe — being significantly less likely to wear a mask when in public than individuals in Japan (the referent country). Individuals were more likely to wear a mask in later months (May: OR 1.71, 95% CI [1.69, 1.75]; June: OR 1.95, 95% CI [1.92, 1.99]; July: OR 2.01, 95% CI [1.97, 2.05]; August: OR 2.60, 95% CI [2.55, 2.65]; September: OR 2.74, 95% CI [2.69, 2.80]; October: OR 3.40, 95% CI [3.32, 3.48]). (**Table 3**).

Demographic, behavioral, and policy-related factors were significantly associated with wearing a face mask in public, even after controlling for time and country fixed effects. Of the demographic factors, female gender (OR 1.70, 95% CI [1.69, 1.71]), living in an urban area (OR 1.40, 95% CI [1.39, 1.41]), and older age (age 25-34: OR 1.22, 95% CI [1.20, 1.23]; age 35-44: OR 1.34, 95% CI [1.32, 1.36]; age 45-54: OR 1.43, 95% CI [1.41, 1.45]; age 55-64: OR 1.42, 95% CI [1.40, 1.44]); age 65+: OR 1.47, 95% CI [1.45, 1.49]) were positively associated with wearing a face mask.

Of the behavioral factors, going to a market, grocery store, or pharmacy was associated with higher mask use (OR 1.07, 95% CI [1.06, 1.08]), whereas more optional or risky behaviors[22] were associated with lower mask use. More specifically, behaviors associated with lower mask use were attending large public events (OR 0.45, 95% CI [0.44, 0.45]), socializing outside of the home [OR 0.72, 95% CI [0.72, 0.73]), and going to a restaurant, cafe, or shopping center (OR 0.77, 95% CI [0.77, 0.78]). Other significant behavioral factors examined were working outside from home, which was associated with lower mask usage (OR 0.98, 95% CI [0.97, 0.99]), and having been tested for COVID-19, which was associated with higher mask use (OR 1.59, 95% CI [1.57, 1.61]).

Regarding country-level factors, we observed that more strict policies were associated with higher mask usage (OR 1.58, 95% CI [1.57, 1.59]), while lagged COVID-19 cases were (OR 0.93, 95% CI [0.92, 0.94]) associated with less mask-wearing.

In the secondary model, which included data from late-June onwards and the three additional socioeconomic variables (financial worry, years of education, and employment status in the last seven days), the aforementioned demographic, behavioral, and policy-related factors remained significantly associated with face mask usage. The three additional socioeconomic variables were significantly associated with mask-wearing: higher years of education was associated with higher use (OR 1.07, 95% CI [1.07, 1.08]) while financial worry and working in the last seven days were associated with lower use (financial worry: OR 0.88, 95% CI [0.87, 0.89]; being employed in the last seven days: OR 0.98, 95% CI [0.97, 0.99]) (**Table 3**).

**Figure 2** depicts the predicted probabilities of wearing a face mask for a few covariates. The results demonstrate that overall, the probabilities of mask-wearing increased over time from April until November but the extent to which the probabilities increased over time varied considerably depending on country (ranging from ~1% increase in Sweden to 50% increase in the United Kingdom; **Figure 2A**). The probability of mask-wearing was also higher among individuals who identify as females (**Figure 2B**) or are living in cities (**Figure 2C**), while it was lower among those who have gone out to a restaurant/shopping center (**Figure 2D**), socialized outside of the household (**Figure 2E**), or attended a large public event (**Figure 2F**). The probabilities varied depending on the country.

## DISCUSSION

In this multi-national sample of over 13 million adults from 38 countries, we found that mask usage has evolved differently across countries during the COVID-19 pandemic. In 13 countries, mask usage prevalence stayed at 70% or higher throughout our study period (April through October), while in Denmark, Sweden, and Norway, mask usage has consistently remained below 15%. In most other countries, mask usage was low in April and eventually reached higher levels, although the pace at which this happened varied widely across these countries. A few other countries have shown more irregular trends over time.

Certain demographics (i.e., female gender, older age, higher education, living in an urban area) were associated with higher mask use, while more optional or risky behaviors, such as attending a large public event, socializing outside of the household, and going to a restaurant, cafe, or shopping center in the last 24 hours were associated with lower mask use. Examining the strength of country-level face covering policies suggested that stricter mask-wearing policies were associated with higher mask usage.

To the best of our knowledge, this is the first study that describes global longitudinal trends of face mask-wearing in a public setting using large nationally representative samples of this scale. Past studies have been limited by narrow geographical coverage and time windows, and most included smaller and non-random samples, with which it is difficult to estimate population representative trends of mask use.[6,8,9] Consequently, it is challenging to compare and contrast our mask use estimates or trends to those from other data sources; however, we found that the trends of mask use for some countries in our study (ex: France, Germany, United Kingdom, and Sweden) are similar to reports provided by other online survey platforms such as YouGov's COVID-19 Public Monitor.[25]

There was considerable heterogeneity in mask use across countries, and some cross-country differences were statistically significant even after adjusting for individual- and country-level factors, such as time-varying mask-wearing policy stringency. These differences in face mask usage across countries suggest that there may be unobserved underlying cultural phenomena across countries that contribute to the adoption of mask-wearing. Pre-existing social norms related to mask-wearing within countries should be taken into consideration when shaping mask-related policy guidelines.

Our findings that certain demographic factors (older age, female gender, and higher education levels) are associated with mask use corroborate findings from past studies reporting that age, gender, and education are significant predictors of face mask usage in the context of other outbreaks, such as SARS-Cov-1 and H1N1.[10–14] One study conducted in Australia reported that those living in rural areas as opposed to urban areas are more likely to wear a face mask;[11] however, in this current, global study, we observed that those living in urban areas are more likely to wear a mask. Notably, the previous study was conducted in the context of an anticipated outbreak scenario, not during an actual pandemic. Taken together, these findings provide a better understanding of who are more or less likely to wear face masks in public during an outbreak and suggest that public health messaging should better target individuals who do not wear face masks in public as frequently.

Interestingly, we found that social behaviors were differentially associated with wearing a mask in that social behaviors that may be deemed more optional and risky in the context of the current pandemic[26] were associated with lower face mask usage, whereas other behaviors that take place outdoors but may be less optional were associated with higher mask use. For example, going out to a large public event, restaurant, cafe, shopping center, or socializing outside of the household in the last 24 hours were associated with lower mask use, whereas going to a market, grocery store, or pharmacy was associated with higher mask use. These results suggest that those who engage in risky social activities during the pandemic are also less likely to wear a mask, and highlight a critical target for public health intervention, as this may contribute to higher risks of COVID-19 spread.[27] Accordingly, our study found that stricter country-level policies around mask-wearing were associated with higher mask use. These results altogether suggest that policies should specifically highlight or put greater emphasis on wearing a face mask in settings where individuals are less likely to do so.

The study has some limitations. First, years of education, financial worry, and employment status were not collected throughout the full field collection period, even though these may be important covariates to examine in association with mask use. To address this, however, we fit a secondary model that included these variables during the narrower time period and found that the results for most associations remained very similar. Second, given our non-experimental study design, we cannot infer any causation from our findings.

Despite the limitations, there are many strengths to this study. This analysis leveraged the largest ongoing, representative data collection related to COVID-19, which allowed us to examine and compare the trends across many countries and include a long time period spanning seven months. We also simultaneously examined individual- and country-level characteristics in our models.

In summary, our study demonstrates various sociodemographic factors, such as older age, female gender, higher education, and urbanicity, are associated with higher face mask usage, while more risky social behaviors, such as going out to a large public event, restaurant, shopping center, and socializing outside of the household are associated with lower mask use. In addition, stronger face mask-related policies are associated with higher mask usage. Taken together, our findings have important implications for health prevention policies and messaging in the context of the ongoing and future public health emergencies, as they highlight the importance of better targeting specific populations and behaviors when designing policies and messaging campaigns.

## DATA SHARING

All data used in this study and from the COVID-19 Symptom Survey are available and publicly available on the University of Maryland's website (https://covidmap.umd.edu/api.html). This website also has the survey questionnaire and detailed documentation of the data that are aggregated and uploaded daily. Individual-level data are available for researchers upon request. For instructions, please visit the website at https://covidmap.umd.edu/.

Other data used in this study, such as those from Johns Hopkins University's COVID-19 data repository and University of Oxford's COVID-19 dataset are also publicly available online.

**ACKNOWLEDGEMENTS**

We are grateful for the support from Kathleen Stewart, Junchuan Fan, Brian Kim in creating the data pipeline, Stanley Presser Alyssa Bilinski, and Josh Salomon, in supporting the questionnaire development. The following Facebook employees were also heavily involved in this project and we are extremely thankful for their support: Tali Alterman Barash, Neta Barkay, Roee Eliat, Andres Garcia, Tal Galili, Andi Gros, Daniel Haimovich, Ahmed Isa, Alex Kaess, Faisal Karim, Ofir Eretz Kedosha, Shelly Matskel, Roee Melamed, Amey Patankar, Irit Rutenberg, Tal Salmona, Tal Sarig, David Vannette. At the University of Maryland we also thank Katie McKeon and Joseph M. Smith for their fast turnarounds on important legal and ethical steps.

**Declaration of interests**

Frauke Kreuter consulted with Facebook from August 2018 – June 2020. Esther Kim, Sarah LaRocca, and Katherine Morris are employed at Facebook and assisted with the interpretation of the results and editing of the manuscript.

## FIGURE LEGENDS

**Figure 1:** Weighted self-reported weekly mask usage prevalence by country, grouped by A) countries with consistently high face mask usage, B) countries that transitioned from low to high face mask usage, C) countries that had consistently low face mask usage, D) countries that showed irregular trends over time

Panel 1A) Mask usage for countries with consistently high face mask usage
Panel 1B) Mask usage for countries that transitioned from low face mask usage to high usage
Panel 1C) Mask usage for countries that had consistently low face mask usage
Panel 1D) Mask usage for countries that showed irregular trends over time

**Figure 2**: Predicted probability of face mask usage by individual characteristics for selected countries given various categories of A) month, B) gender, C) urbanicity, D) having gone to a restaurant, café or shopping center, E) having socialized outside of the household, and D) having attended a large public event

Panel 2A) By month (04=April, 10=October)
Panel 2B) By gender (1=Female, 0=Not Female)
Panel 2C) By urbanicity (1=Urban, 0=Not Urban)
Panel 2D) By going out to a restaurant, café, or shopping center (1=Yes, 0=No)
Panel 2E) By socializing outside the household (1=Yes, 0=No)
Panel 2F) By attending a large public event (1=Yes, 0=No)

**Table 1**: List of countries included in the analysis

| Country | Total responses | Complete responses[a] | Analytic sample[b] |
|---|---|---|---|
| | 18,730,575 (100%) | 14,552,118 (77.69%) | 13,723,810 (73.27%) |
| Argentina | 659,009 (3.52%) | 484,352 (2.59%) | 435,134 (2.32%) |
| Australia | 325,885 (1.74%) | 266,113 (1.42%) | 254,095 (1.36%) |
| Austria | 138,777 (0.74%) | 113,097 (0.6%) | 111,133 (0.59%) |
| Belgium | 142,999 (0.76%) | 101,548 (0.54%) | 97,550 (0.52%) |
| Brazil | 2,322,508 (12.4%) | 1,788,903 (9.55%) | 1,700,210 (9.08%) |
| Bulgaria | 96,459 (0.51%) | 75,085 (0.4%) | 72,191 (0.39%) |
| Canada | 417,071 (2.23%) | 346,718 (1.85%) | 329,517 (1.76%) |
| Chile | 324,447 (1.73%) | 256,195 (1.37%) | 229,449 (1.22%) |
| Colombia | 574,169 (3.07%) | 449,043 (2.4%) | 394,540 (2.11%) |
| Costa Rica | 167,986 (0.9%) | 131,144 (0.7%) | 115,907 (0.62%) |
| Czech Republic | 213,108 (1.14%) | 168,216 (0.9%) | 162,533 (0.87%) |
| Denmark | 306,917 (1.64%) | 257,938 (1.38%) | 254,406 (1.36%) |
| Finland | 157,593 (0.84%) | 133,380 (0.71%) | 125,433 (0.67%) |
| France | 708,994 (3.79%) | 459,218 (2.54%) | 442,412 (2.36%) |
| Germany | 763,760 (4.08%) | 628,053 (3.35%) | 619,066 (3.31%) |
| Greece | 197,813 (1.06%) | 163,753 (0.87%) | 154,268 (0.82%) |
| Hungary | 320,668 (1.71%) | 255,230 (1.36%) | 242,901 (1.3%) |
| India | 1,083,384 (5.78%) | 728,852 (3.89%) | 642,297 (3.43%) |
| Indonesia | 547,797 (2.92%) | 398,395 (2.13%) | 370,107 (1.98%) |
| Ireland | 163,006 (0.87%) | 131,488 (0.7%) | 125,645 (0.67%) |
| Israel | 193,693 (1.03%) | 156,551 (0.84%) | 150,858 (0.81%) |
| Italy | 989,919 (5.29%) | 796,122 (4.25%) | 775,840 (4.14%) |
| Japan | 1,418,201 (7.57%) | 1,178,538 (6.29%) | 1,163,828 (6.21%) |
| Mexico | 1,831,010 (9.78%) | 1,425,019 (7.61%) | 1,302,477 (6.95%) |
| Netherlands | 355,421 (1.9%) | 294,844 (1.57%) | 265,162 (1.42%) |
| New Zealand | 125,601 (0.67%) | 101,695 (0.54%) | 96,472 (0.52%) |
| Norway | 205,460 (1.1%) | 171,999 (0.92%) | 157,285 (0.84%) |
| Poland | 388,553 (2.07%) | 265,143 (1.42%) | 257,927 (1.38%) |
| Portugal | 353,606 (1.89%) | 247,065 (1.32%) | 238,697 (1.27%) |
| Romania | 385,949 (2.06%) | 308,116 (1.64%) | 293,426 (1.57%) |
| Russia | 273,870 (1.46%) | 214,795 (1.15%) | 201,447 (1.08%) |
| Slovenia | 43,665 (0.23%) | 35,613 (0.19%) | 34,606 (0.18%) |
| South Africa | 235,188 (1.26%) | 189,818 (1.01%) | 179,337 (0.96%) |
| Spain | 659,951 (3.52%) | 518,427 (2.77%) | 503,994 (2.69%) |
| Sweden | 484,930 (2.59%) | 409,192 (2.18%) | 385,530 (2.06%) |
| Switzerland | 148,638 (0.79%) | 115,230 (0.62%) | 112,087 (0.6%) |
| Turkey | 480,588 (2.57%) | 359,907 (1.92%) | 336,594 (1.8%) |
| United Kingdom | 523,982 (2.8%) | 427,323 (2.28%) | 389,449 (2.08%) |

**Table 2**: Weighted distribution of respondent characteristics among 13,723,810 respondents who reported being in public in the past 7 days and provided complete responses

| | Overall (Unweighted *N* = 13,723,810) | Mask usage = 1[a] (Unweighted *N* = 10,610,836) | Mask usage = 0[b] (Unweighted *N* = 3,112,974) |
|---|---|---|---|
| Sex | % | % | % |
| Female | 43.15 | 43.96 | 39.03 |
| Male | 56.64 | 55.85 | 60.66 |
| Other | 0.20 | 0.18 | 0.30 |
| Age | % | % | % |
| 18 – 34 | 14.68 | 14.71 | 14.52 |
| 25 – 34 | 26.29 | 26.97 | 22.83 |
| 35 – 44 | 18.76 | 18.84 | 18.34 |
| 45 – 54 | 18.3 | 18.24 | 18.61 |
| 55 – 64 | 10.70 | 10.40 | 12.23 |
| >= 65 | 11.26 | 10.84 | 13.44 |
| Current location | % | % | % |
| Urban | 46.37 | 44.33 | 56.75 |
| Non-urban | 53.63 | 55.67 | 43.25 |
| Gone to work outside in the last 24 hours | % | % | % |
| Yes | 36.23 | 35.63 | 39.29 |
| No | 63.76 | 64.36 | 60.71 |
| Gone to a market, grocery store, or pharmacy in the last 24 hours | % | % | % |
| Yes | 65.79 | 65.24 | 68.58 |
| No | 35.96 | 34.76 | 31.42 |
| Gone to a restaurant, café, or shopping center in the last 24 hours | % | % | % |
| Yes | 25.36 | 23.84 | 33.11 |
| No | 74.64 | 76.15 | 66.89 |
| Spent time with a non-same household member in the last 24 hours | % | % | % |

| | | | | | | | | | |
|---|---|---|---|---|---|---|---|---|---|
| Yes | | 43.16 | | | 40.70 | | | 55.68 | |
| No | | 56.84 | | | 59.29 | | | 44.32 | |
| Attended a public event with more than 10 people in the last 24 hours | | % | | | % | | | % | |
| Yes | | 10.31 | | | 9.65 | | | 17.62 | |
| No | | 89.05 | | | 90.35 | | | 82.38 | |
| Tested for COVID-19 | | % | | | % | | | % | |
| Yes | | 12.94 | | | 13.36 | | | 10.76 | |
| No | | 87.06 | | | 86.64 | | | 89.22 | |
| Worried about household finances in the next month* | | % | | | % | | | % | |
| Yes | | 21.01 | | | 21.35 | | | 17.68 | |
| No | | 78.99 | | | 78.65 | | | 82.32 | |
| Worked for pay in the last 7 days* | | % | | | % | | | % | |
| Yes | | 53.10 | | | 53.84 | | | 58.79 | |
| No | | 46.90 | | | 46.16 | | | 41.21 | |
| Years of education* | Q1 | Q2 | Q3 | Q1 | Q2 | Q3 | Q1 | Q2 | Q3 |
| | 10 | 14 | 17 | 11 | 15 | 17 | 9 | 13 | 16 |

**Table 3**: Weighted logistic regression results on the associations of various individual-level and country-level variables with face mask use[1]

| | **Primary model: From April 22 until October 31** | | **Secondary model: From June 27 until October 31** | |
|---|---|---|---|---|
| | **Odds ratio (95% CI)** | **P** | **Odds ratio (95% CI)** | **P** |
| **Time (month)** | | | | |
| **April** | *ref* | | Not available | |
| **May** | 1.72 (1.69, 1.75) | <0.001 | Not available | |
| **June** | 1.95 (1.92, 1.99) | <0.001 | *ref* | |
| **July** | 2.01 (1.97, 2.05) | <0.001 | 1.14 (1.11, 1.17) | <0.001 |
| **August** | 2.60 (2.55, 2.65) | <0.001 | 1.45 (1.42, 1.48) | <0.001 |
| **September** | 2.74 (2.69, 2.80) | <0.001 | 1.51 (1.48, 1.55) | <0.001 |
| **October** | 3.40 (3.32, 3.48) | <0.001 | 1.90 (1.85, 1.96) | <0.001 |
| **COVID-19 test ever taken** | | | | |
| **No** | *ref* | | *ref* | |
| **Yes** | 1.59 (1.57, 1.61) | <0.001 | 1.55 (1.53, 1.58) | <0.001 |

[1] Dependent variable: mask usage (binary), type: Analysis of complex survey design; link function: logit. Model includes fixed effects by country (not shown in Table). Pseudo $R^2 = 0.27$ for the primary model; Pseudo $R^2 = 0.26$ for secondary model.

| | | | | |
|---|---|---|---|---|
| **Age** | | | | |
| **18-24 years** | *ref* | | *ref* | |
| **25-34 years** | 1.22 (1.20, 1.23) | <0.001 | 1.16 (1.14, 1.18) | <0.001 |
| **35-44 years** | 1.34 (1.32, 1.36) | <0.001 | 1.23 (1.21, 1.26) | <0.001 |
| **45-54 years** | 1.43 (1.41, 1.45) | <0.001 | 1.32 (1.29, 1.34) | <0.001 |
| **55-64 years** | 1.42 (1.40, 1.44) | <0.001 | 1.28 (1.25, 1.30) | <0.001 |
| **65+ years** | 1.47 (1.45, 1.50) | <0.001 | 1.27 (1.24, 1.29) | <0.001 |
| **Gender** | | | | |
| **Male/other** | *ref* | | *ref* | |
| **Female** | 1.70 (1.69, 1.71) | <0.001 | 1.75 (1.73, 1.77) | <0.001 |
| **Living in an urban area** | | | | |
| **No** | *ref* | | *ref* | |
| **Yes** | 1.40 (1.39, 1.41) | <0.001 | 1.43 (1.41, 1.44) | <0.001 |
| **Gone out to work outside in the last 24 hours** | | | | |
| **No** | *ref* | | *ref* | |
| **Yes** | 0.98 (0.98, 0.99) | <0.001 | 0.98 (0.97, 0.99) | <0.001 |
| **Gone out to a market, grocery store or pharmacy in the last 24 hours** | | | | |
| **No** | *ref* | | *ref* | |
| **Yes** | 1.07 (1.06, 1.08) | <0.001 | 1.01 (1.09, 1.11) | <0.001 |
| **Gone out to a restaurant, café, or shopping center in the last 24 hours** | | | | |
| **No** | *ref* | | *ref* | |
| **Yes** | 0.77 (0.77, 0.78) | <0.001 | 0.76 (0.75, 0.77) | <0.001 |
| **Spent time with someone outside their household in the last 24 hours** | | | | |
| **No** | *ref* | | *ref* | |
| **Yes** | 0.72 (0.72, 0.73) | <0.001 | 0.70 (0.69, 0.71) | <0.001 |
| **Attended a public event with more than 10 people in the last 24 hours** | | | | |
| **No** | *ref* | | *ref* | |
| **Yes** | 0.45 (0.44, 0.45) | <0.001 | 0.46 (0.46, 0.47) | <0.001 |
| **Mask policy stringency score** | | | | |
| **Per 1 standard deviation** | 1.58 (1.58, 1.59) | <0.001 | 1.50 (1.48, 1.51) | <0.001 |
| **Seven day lagged COVID-19 cases** | | | | |
| **Per 1 standard deviation** | 0.93 (0.93, 0.94) | <0.001 | 0.98 (0.98, 0.99) | <0.001 |
| **Worked in the last 7 days** | | | | |
| **No** | Not available | | *ref* | |
| **Yes** | Not available | | 0.98 (0.97, 0.99) | <0.001 |
| **Worried about household finances** | | | | |
| **No** | Not available | | *ref* | |
| **Yes** | Not available | | 0.88 (0.87, 0.89) | <0.001 |
| **Years of education** | | | | |

| | | | |
|---|---|---|---|
| **Per 1 year** | Not available | 1.07 (1.07, 1.08) | <0.001 |

**Figure 1:** Weighted self-reported weekly mask usage prevalence by country[2], grouped by A) countries with consistently high face mask usage, B) countries that transitioned from low to high face mask usage, C) countries that had consistently low face mask usage, D) countries that showed irregular trends over time

**Panel 1A) Mask usage for countries with consistently high face mask usage**

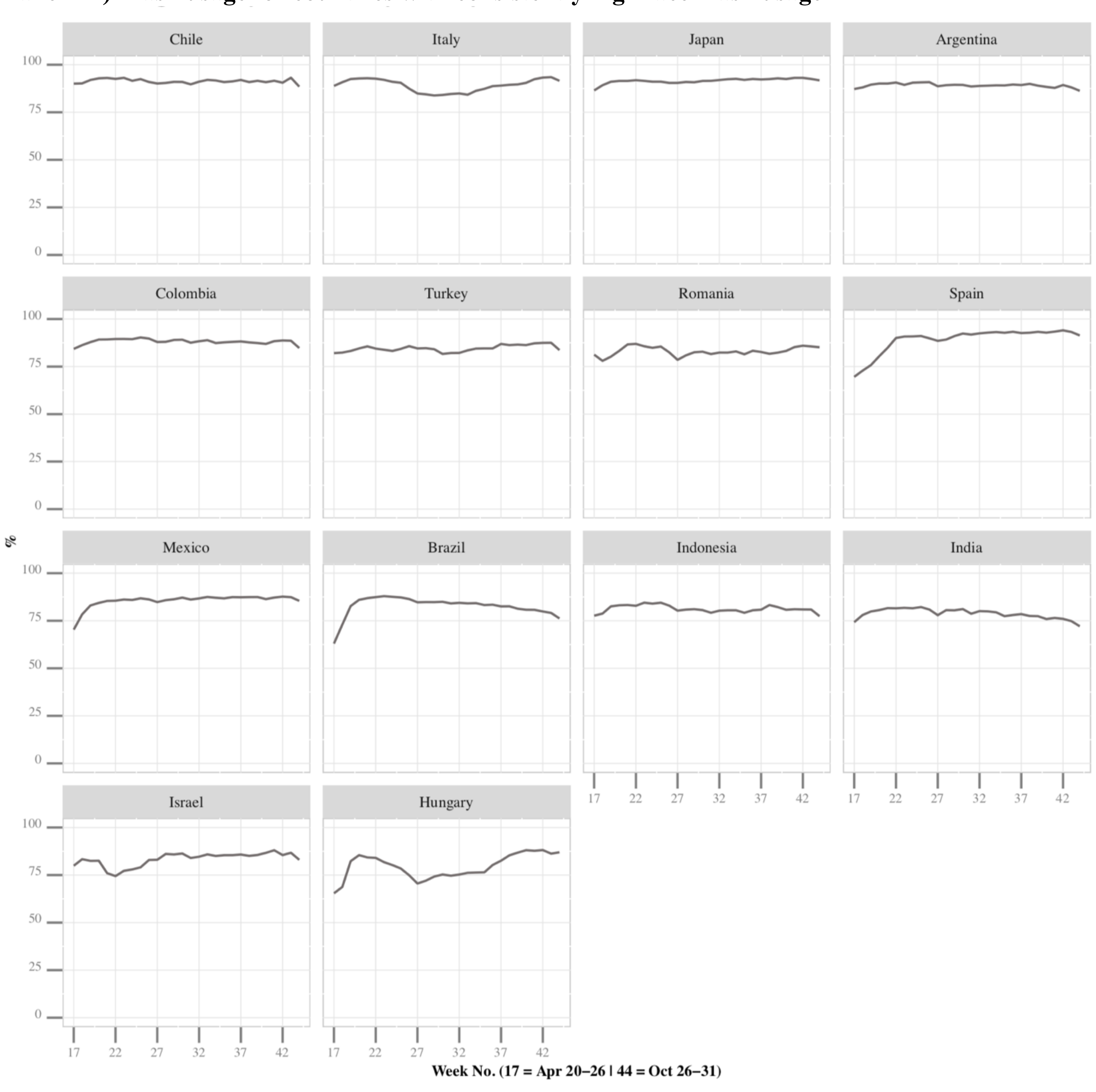

[2] Weights adjust each country sample to their corresponding national population.

**Figure 1:** Continued (Part 2)

**Panel 1B) Mask usage for countries that transitioned from low face mask usage to high usage**

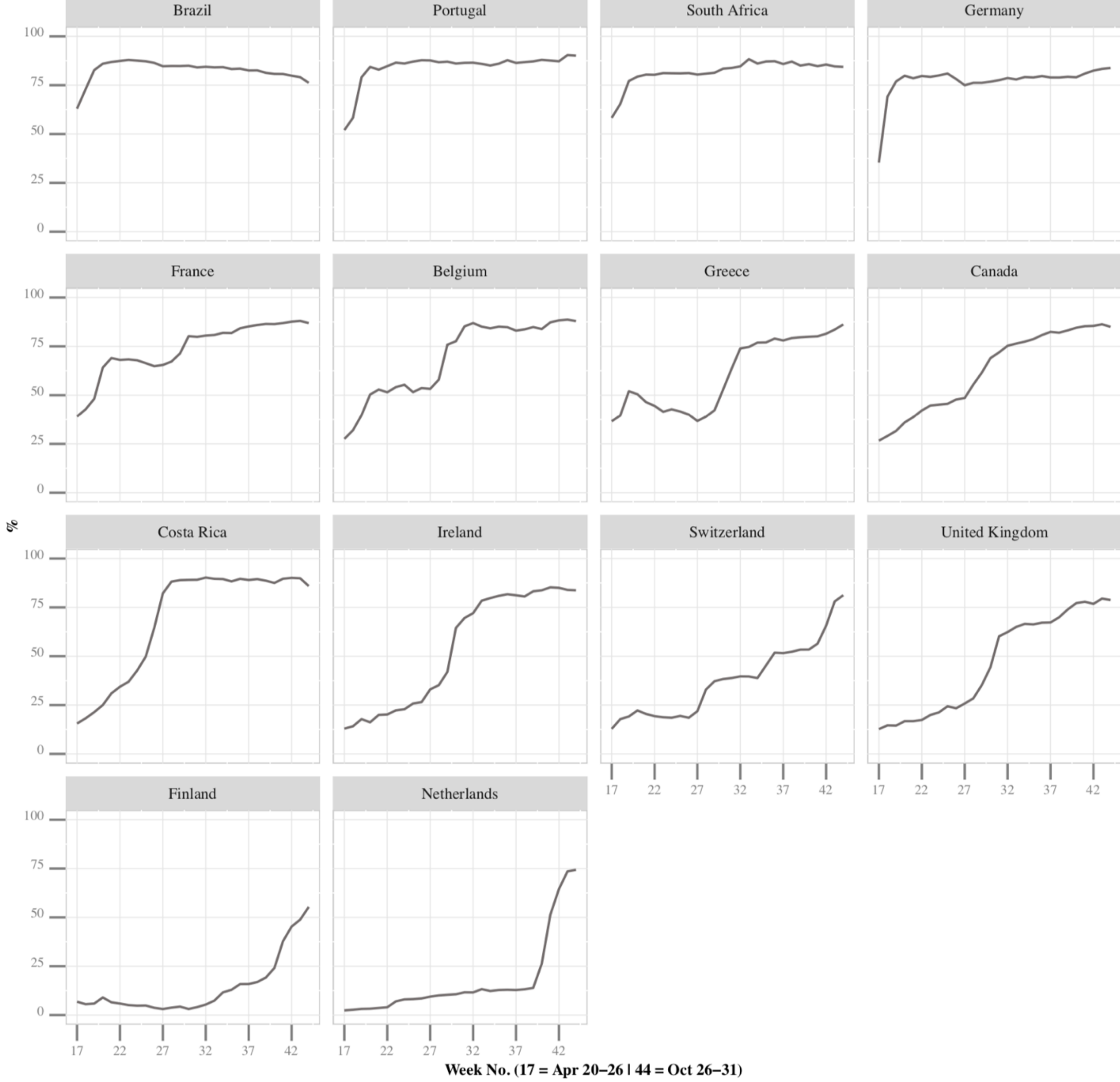

**Figure 2:** Continued (Part 3)

**Panel 1C) Mask usage for countries that had consistently low face mask usage**

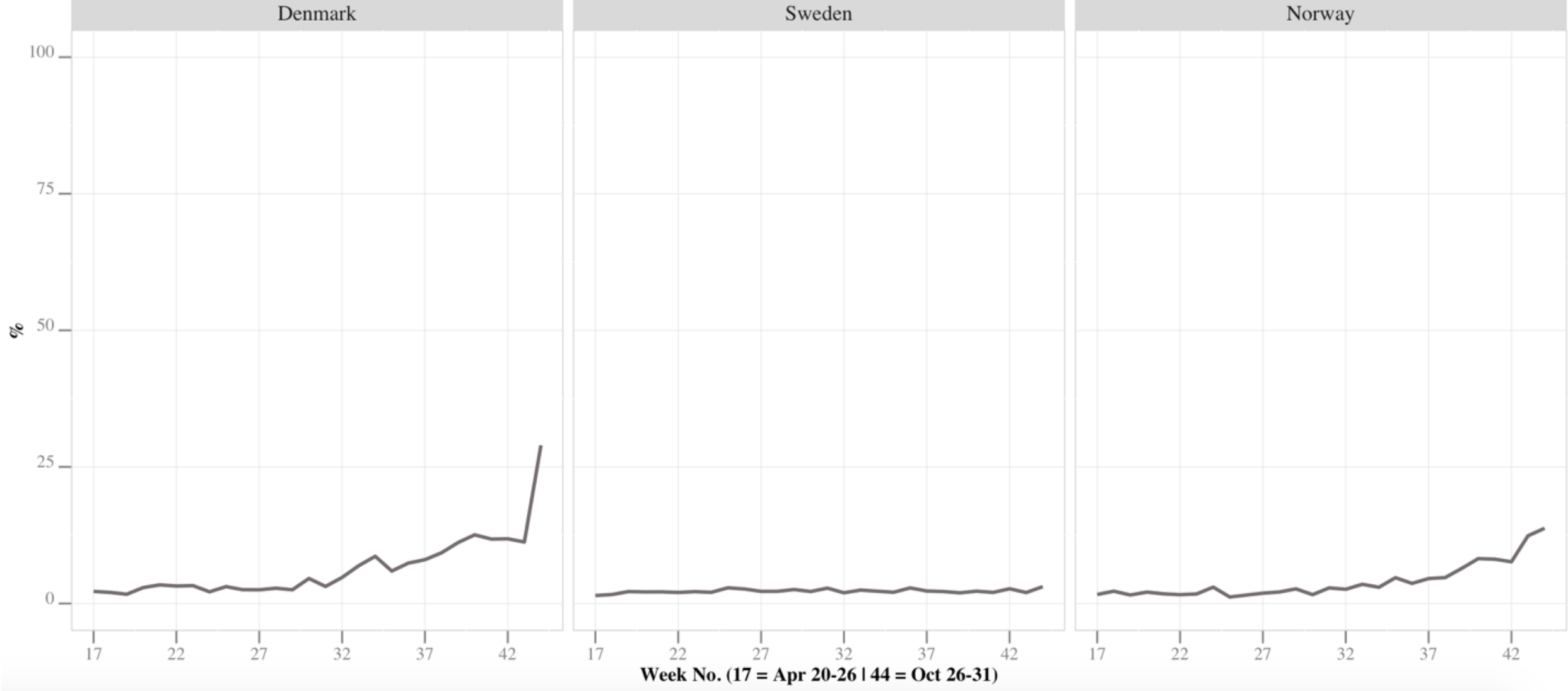

**Figure 1**: Continued (Part 4)

**Panel 1D) Mask usage for countries that showed irregular trends over time**

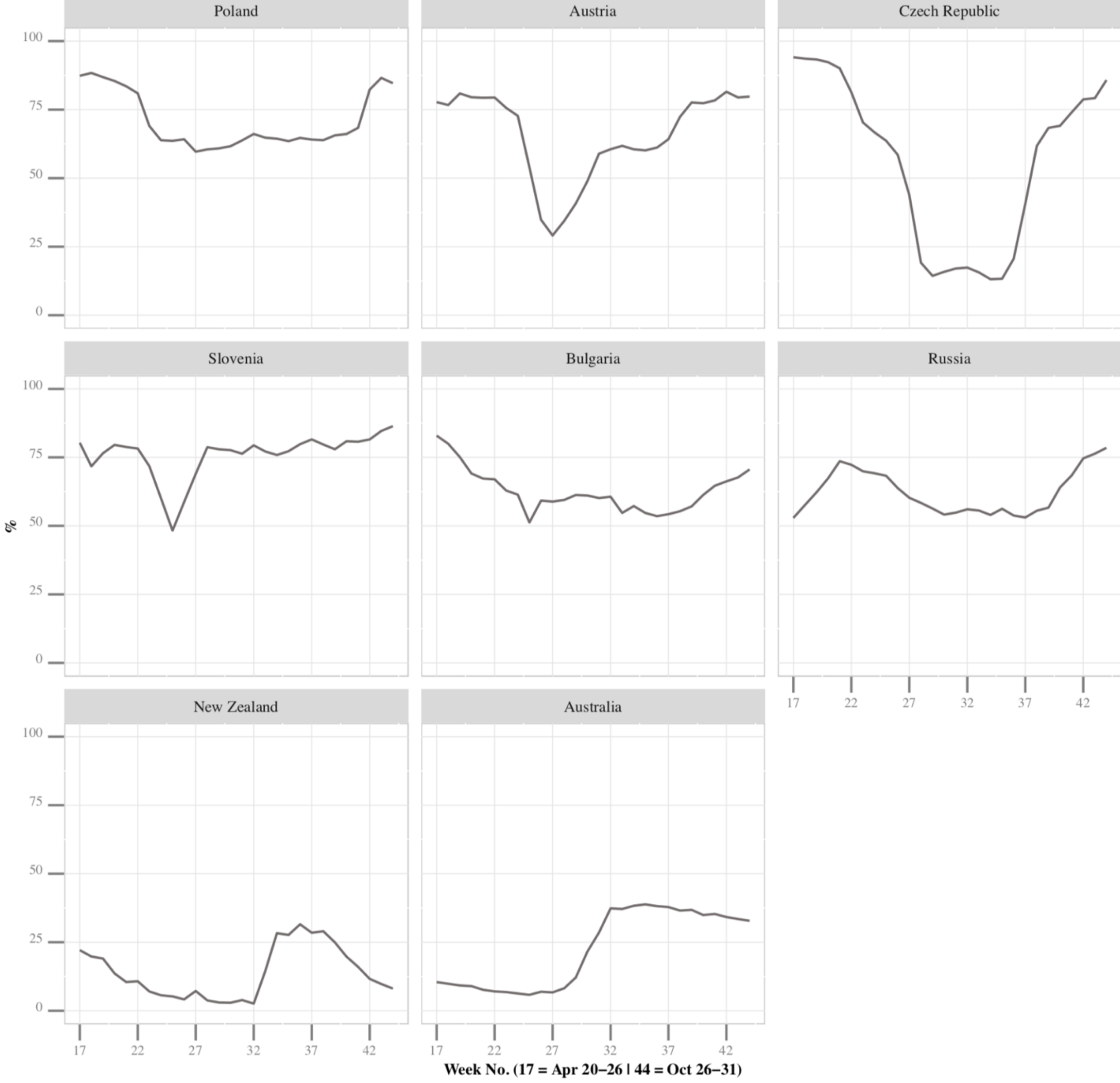

**Figure 2**: Predicted probability of face mask usage by individual characteristics for selected countries given various categories of A) month, B) gender, C) urbanicity, D) having gone to a restaurant, café or shopping center, E) having socialized outside of the household, and D) having attended a large public event

**Panel 2A) By month (04=April, 10=October)**

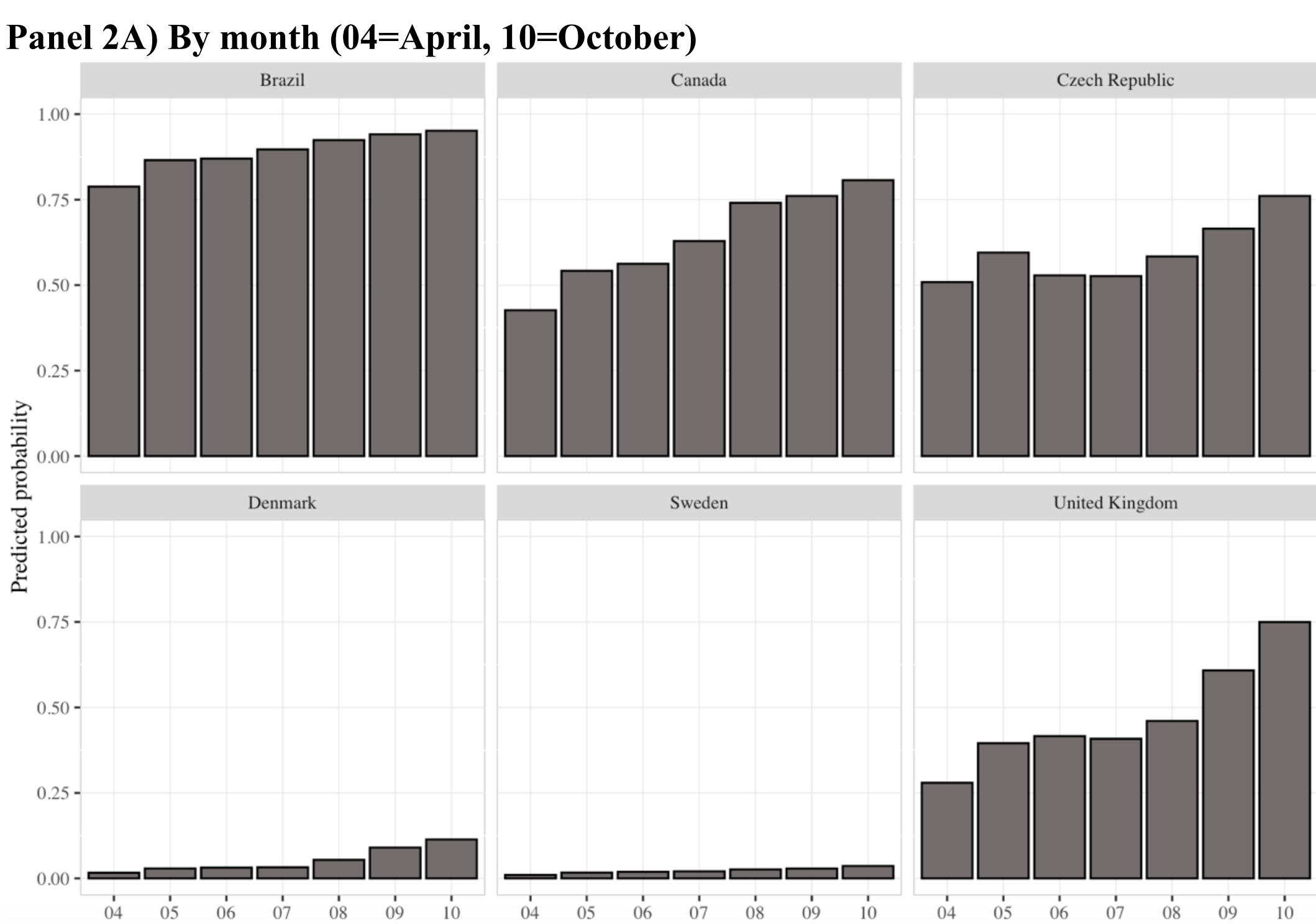

**Panel 2B) By gender (1=Female, 0=Not Female)**

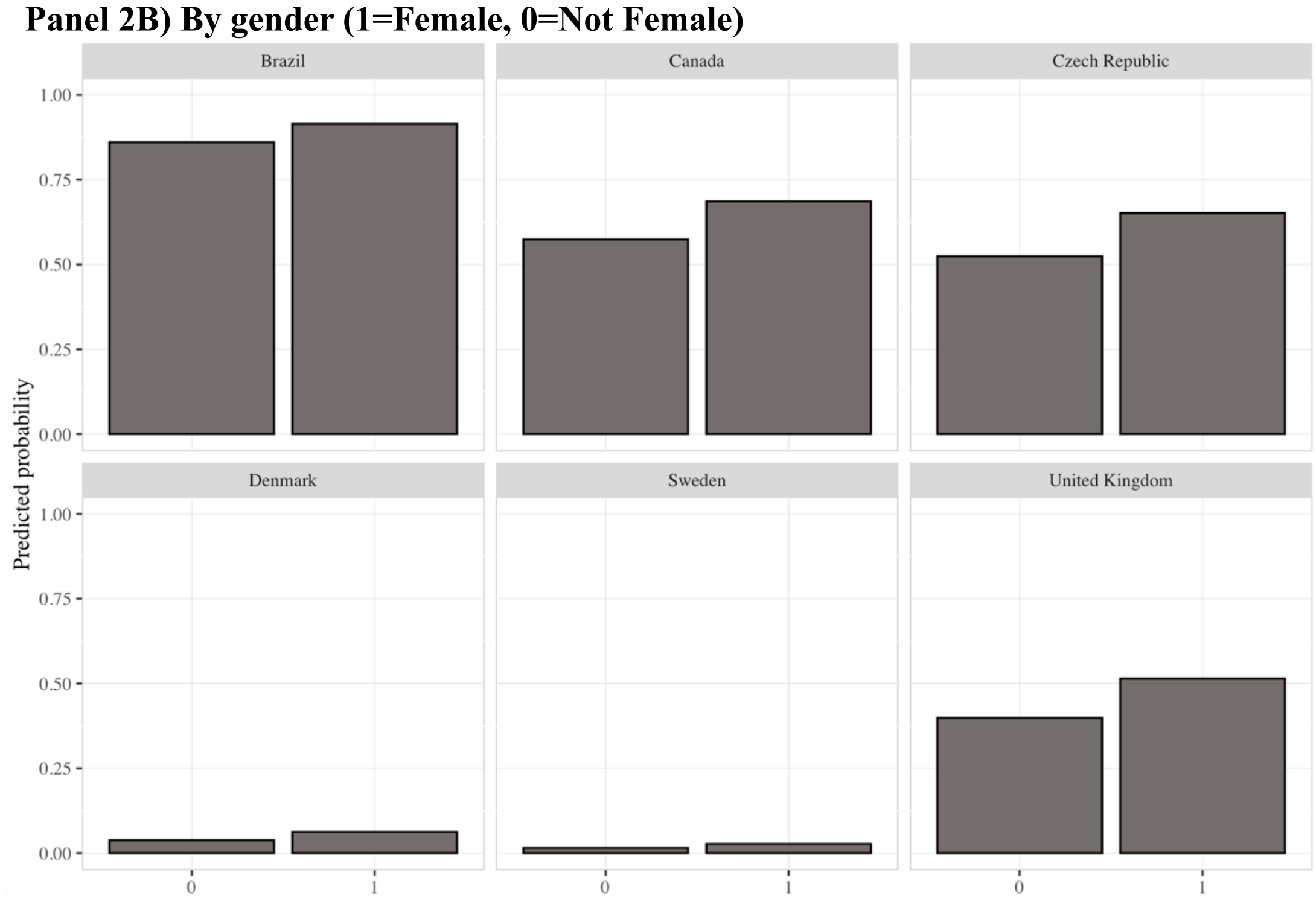

**Figure 2**: Continued (Part 2)

**Panel 2C) By urbanicity (1=Urban, 0=Not Urban)**

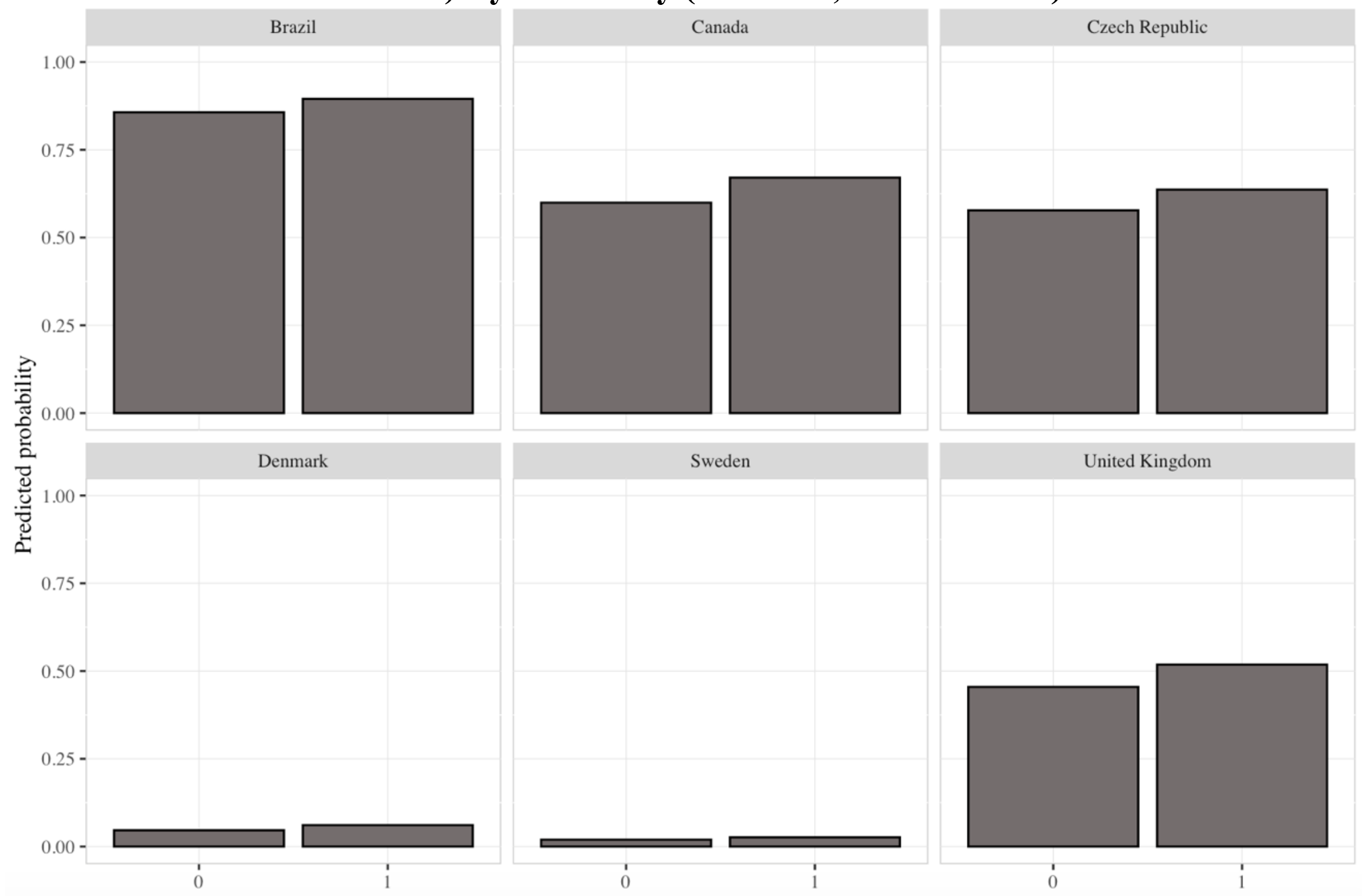

**Panel 2D) By going out to a restaurant, café, or shopping center (1=Yes, 0=No)**

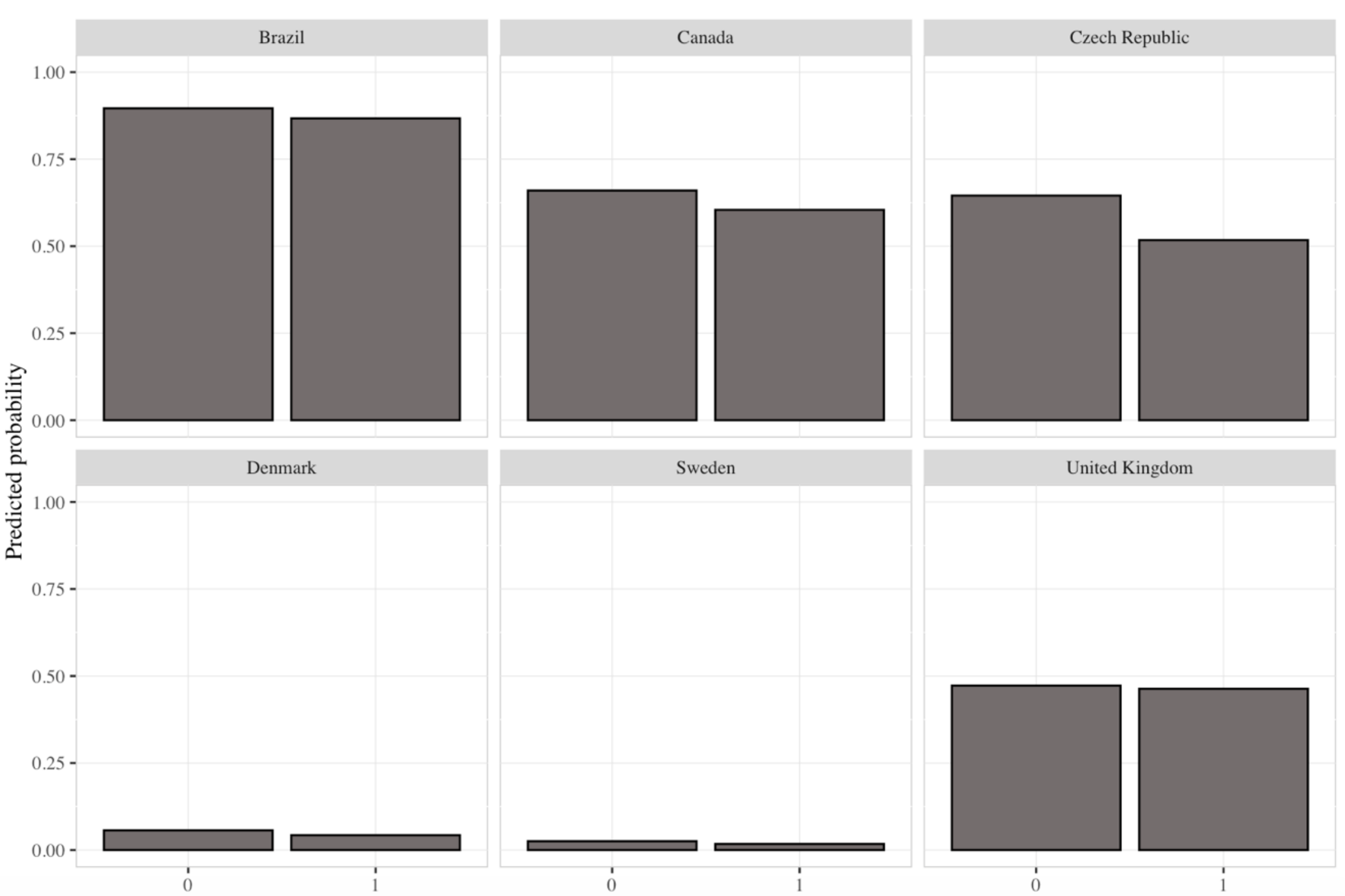

**Figure 2**: Continued (Part 3)

**Panel 2E) By socializing outside the household (1=Yes, 0=No)**

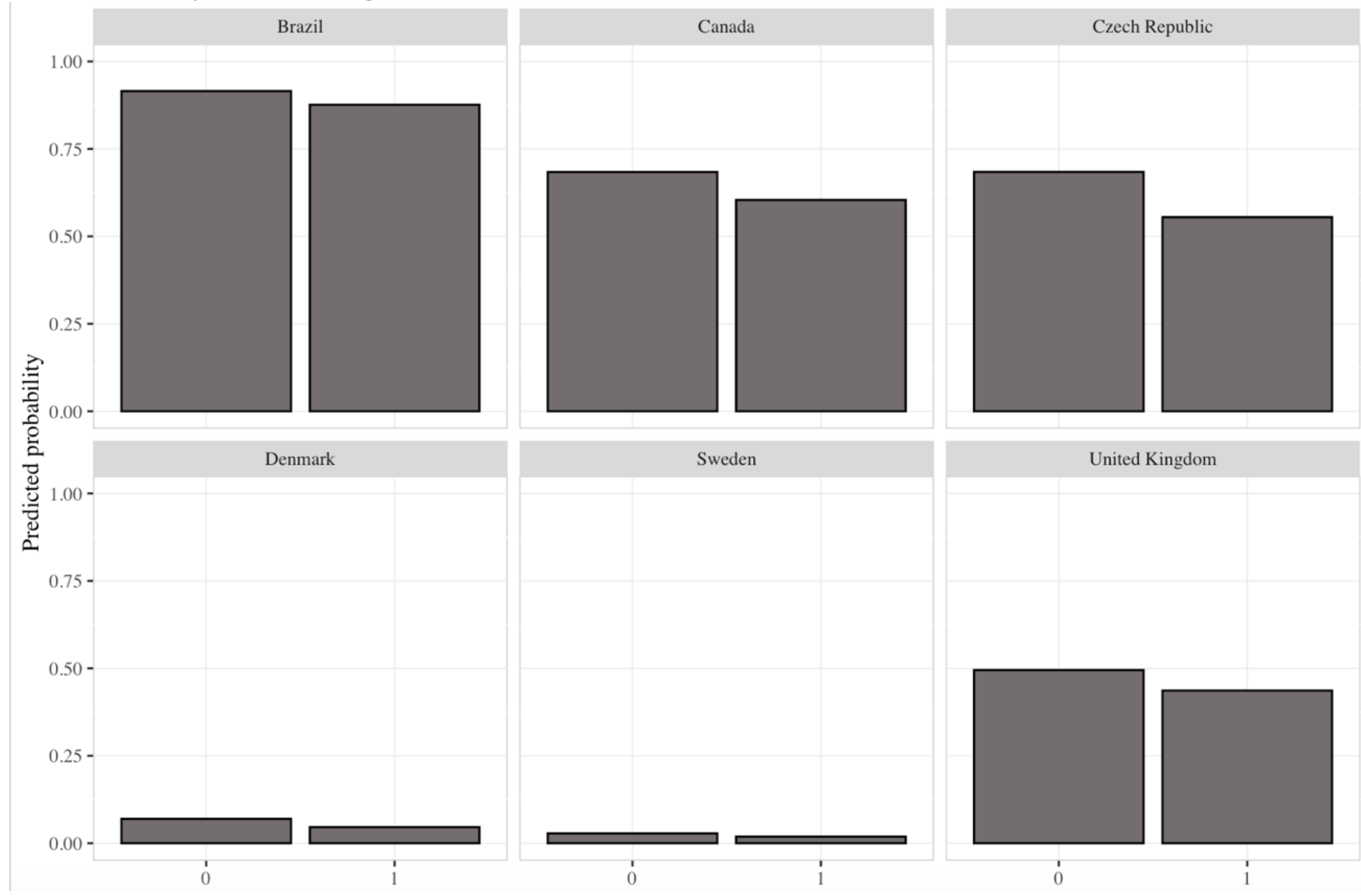

**Panel 2F) By attending a large public event (1=Yes, 0=No)**

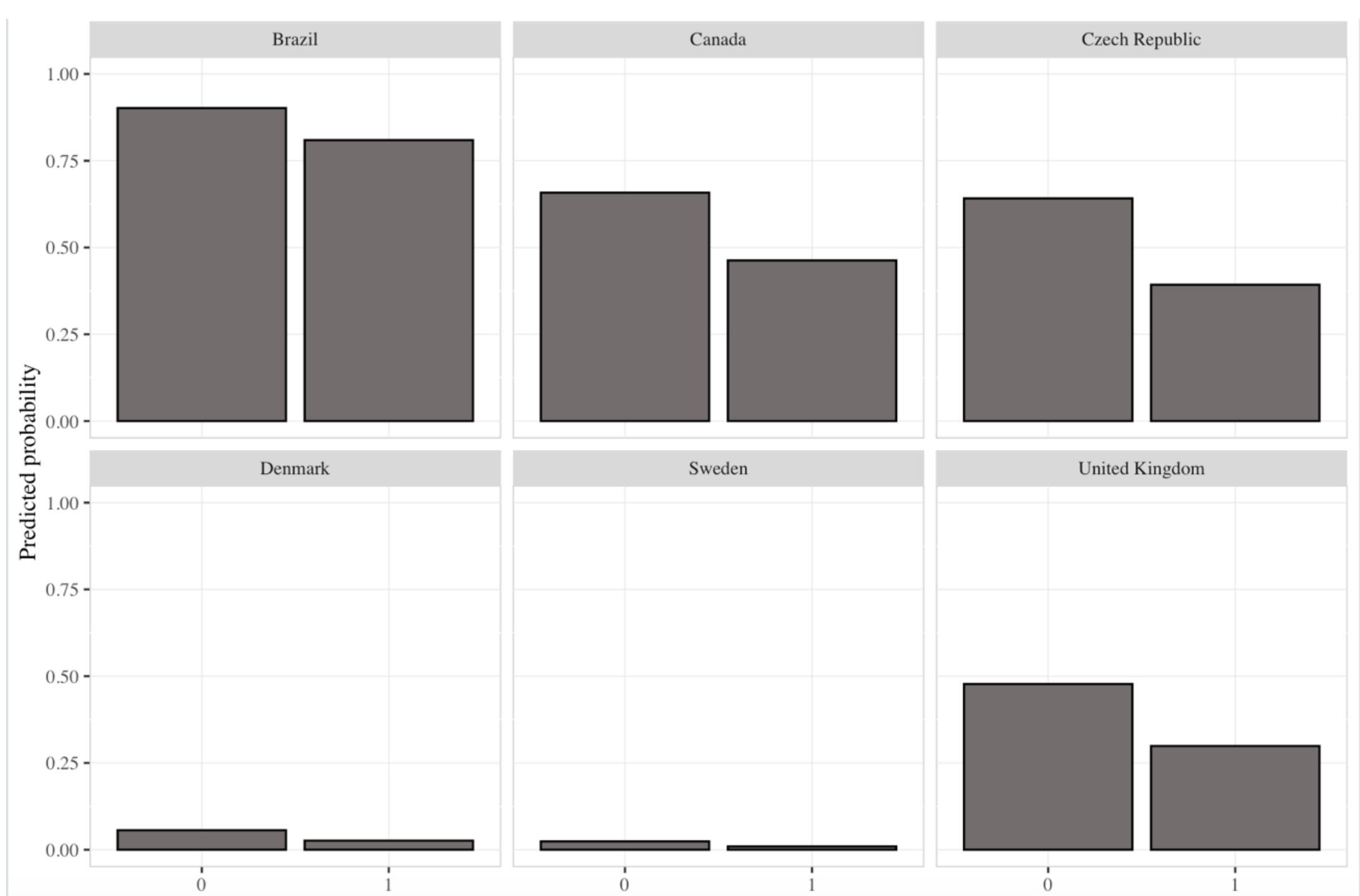

## SUPPLEMENTARY MATERIALS

**Supplementary Table 1**: List of countries, territories, and languages supported by the COVID-19 Symptom Survey

| Countries or territories | Languages |
|---|---|
| Andorra; United Arab Emirates; Afghanistan; Antigua and Barbuda; Anguilla; Albania; Armenia; Netherland Antilles; Angola; Antarctica; Argentina; American Samoa; Austria; Australia; Aruba; Azerbaijan; Bosnia and Herzegovina; Barbados; Bangladesh; Belgium; Burkina Faso; Bulgaria; Bahrain; Burundi; Benin; Saint Barthélemy; Bermuda; Brunei Darussalam; Bolivia (Plurinational State of); Bonaire, Sint Eustatius and Saba; Brazil; Bahamas; Bhutan; Botswana; Belarus; Belize; Canada; Congo, Democratic Republic of the; Central African Republic; Congo; Switzerland; Côte d'Ivoire; Cook Islands; Chile; Cameroon; China; Colombia; Costa Rica; Cuba; Cabo Verde; Curaçao; Cyprus; Czechia; Germany; Djibouti; Denmark; Dominica; Dominican Republic; Algeria; Ecuador; Estonia; Egypt; Western Sahara; Eritrea; Spain; Ethiopia; Finland; Fiji; Falkland Islands (Malvinas); Micronesia (Federated States of); Faroe Islands; France; Gabon; United Kingdom of Great Britain and Northern Ireland; Grenada; Georgia; French Guiana; Guernsey; Ghana; Gibraltar; Greenland; Gambia; Guinea; Guadeloupe; Equatorial Guinea; Greece; Guatemala; Guam; Guinea-Bissau; Guyana; Hong Kong; Honduras; Croatia; Haiti; Hungary; Indonesia; Ireland; Israel; Isle of Man; India; Iraq; Iran (Islamic Republic of); Iceland; Italy; Jersey; Jamaica; Jordan; Japan; Kenya; Kyrgyzstan; Cambodia; Kiribati; Comoros; Saint Kitts and Nevis; Korea, Republic of; Kuwait; Cayman Islands; Kazakhstan; Lao People's Democratic Republic; Lebanon; Saint Lucia; Liechtenstein; Sri Lanka; Liberia; Lesotho; Lithuania; Luxembourg; Latvia; Libya; Morocco; Monaco; Moldova, Republic of; Montenegro; Saint Martin (French part); Madagascar; Marshall Islands; Macedonia, the former Yugoslav Republic of; Mali; Myanmar; Mongolia; Macao; Northern Mariana Islands; Martinique; Mauritania; Montserrat; Malta; Mauritius; Maldives; Malawi; Mexico; Malaysia; Mozambique; Namibia; New Caledonia; Niger; Norfolk Island; Nigeria; Nicaragua; Netherlands; Norway; Nepal; Nauru; Niue; New Zealand; Oman; Panama; Peru; French Polynesia; Papua New Guinea; Philippines; Pakistan; Poland; Saint Pierre and Miquelon; Puerto Rico; Palestine, State of; Portugal; Palau; Paraguay; Qatar; Réunion; Romania; Serbia; Russian Federation; Rwanda; Saudi Arabia; Solomon Islands; Seychelles; Sudan; Sweden; Singapore; Saint Helena, Ascension and Tristan da Cunha; Slovenia; Svalbard and Jan Mayen; Slovakia; Sierra Leone; San Marino; Senegal; Somalia; Suriname; South Sudan; Sao Tome and Principe; El Salvador; Sint Maarten (Dutch part); Syrian Arab Republic; Eswatini; Turks and Caicos Islands; Chad; Togo; Thailand; Tajikistan; Tokelau; Timor-Leste; Turkmenistan; Tunisia; Tonga; Turkey; Trinidad and Tobago; Tuvalu; Taiwan, Province of China; Tanzania, United Republic of; Ukraine; Uganda; United States of America; Uruguay; Uzbekistan; Holy See; Saint Vincent and the Grenadines; Venezuela (Bolivarian Republic of); Virgin Islands (British); Virgin Islands (U.S.); Viet Nam; Vanuatu; Wallis and Futuna; Samoa; Kosovo; Yemen; Mayotte; South Africa; Zambia; Zimbabwe | Arabic; Azerbaijani; Bulgarian; Bengali; Czech; Cebuano; Danish; German; Greek; English (UK); NA; Spanish (Spain); Spanish; Persian; Finnish; French (Canada); French (France); Gujarati; Hebrew; Hindi; Croatian; Hungarian; Indonesian; Italian; Japanese; Kannada; Korean; Macedonian; Malayalam; Marathi; Malay; Burmese; Norwegian (bokmal); Dutch; Punjabi; Polish; Portuguese (Brazil); Portuguese (Portugal); Romanian; Russian; Slovak; Slovenian; Albanian; Serbian; Swedish; Swahili; Tamil; Telugu; Thai; Filipino; Turkish; Urdu; Vietnamese; Simplified Chinese (China); Traditional Chinese (Hong Kong); Traditional Chinese (Taiwan) |

**Supplementary Table 2**: Weighted distribution of respondent characteristics among 18,730,575 respondents who responded to the survey

| | Overall (Unweighted *N* = 18,730,575) | Mask usage = 1[a] (Unweighted *N* = 13,006,455) | Mask usage = 0[b] (Unweighted *N* = 3,655,440) |
|---|---|---|---|
| Sex | % | % | % |
| Female | 36.37 | 39.57 | 35.50 |
| Male | 45.83 | 49.61 | 55.12 |
| Other | 0.18 | 0.18 | 0.30 |
| Missing | 17.61 | 10.64 | 9.09 |
| Age | % | % | % |
| 18 – 34 | 13.01 | 13.45 | 13.63 |
| 25 – 34 | 21.77 | 23.98 | 20.78 |
| 35 – 44 | 15.07 | 16.77 | 16.64 |
| 45 – 54 | 14.66 | 16.35 | 16.92 |
| 55 – 64 | 8.93 | 9.59 | 11.32 |
| >= 65 | 9.97 | 10.33 | 12.84 |
| Missing | 16.59 | 9.52 | 7.88 |
| Current location | % | % | % |
| Urban | 43.26 | 48.99 | 38.97 |
| Non-urban | 38.36 | 39.45 | 51.34 |
| Missing | 18.38 | 11.56 | 9.69 |
| Gone to work outside in the last 24 hours | % | % | % |
| Yes | 32.33 | 35.40 | 39.31 |
| No | 61.42 | 62.61 | 58.73 |
| Missing | 6.25 | 1.99 | 1.95 |
| Gone to a market, grocery store, or pharmacy in the last 24 hours | % | % | % |
| Yes | 58.33 | 64.38 | 68.01 |
| No | 35.96 | 34.25 | 30.66 |
| Missing | 5.71 | 1.37 | 1.33 |
| Gone to a restaurant, café, or shopping center in the last 24 hours | % | % | % |
| Yes | 22.46 | 23.79 | 33.17 |
| No | 70.73 | 73.64 | 64.39 |
| Missing | 6.82 | 2.57 | 2.45 |
| Spent time with a non-same household member in the last 24 hours | % | % | % |
| Yes | 37.88 | 39.68 | 54.89 |

| No | | 55.91 | | | 58.38 | | | 43.44 | |
|---|---|---|---|---|---|---|---|---|---|
| Missing | | 6.21 | | | 1.94 | | | 1.67 | |
| Attended a public event with more than 10 people in the last 24 hours | | % | | | % | | | % | |
| Yes | | 10.31 | | | 10.06 | | | 18.63 | |
| No | | 83.18 | | | 87.69 | | | 79.33 | |
| Missing | | 6.51 | | | 2.26 | | | 2.04 | |
| Tested for COVID-19 | | % | | | % | | | % | |
| Yes | | 13.72 | | | 14.05 | | | 11.31 | |
| No | | 83.50 | | | 85.03 | | | 87.82 | |
| Missing | | 2.77 | | | 0.93 | | | 0.88 | |
| Worried about household finances in the next month* | | % | | | % | | | % | |
| Yes | | 18.64 | | | 20.72 | | | 17.50 | |
| No | | 68.85 | | | 75.16 | | | 78.57 | |
| Missing | | 12.52 | | | 4.13 | | | 3.93 | |
| Worked for pay in the last 7 days* | | % | | | % | | | % | |
| Yes | | 44.49 | | | 49.46 | | | 54.35 | |
| No | | 41.95 | | | 45.25 | | | 40.70 | |
| Missing | | 13.56 | | | 5.29 | | | 4.94 | |
| Years of education* | Q1 | Q2 | Q3 | Q1 | Q2 | Q3 | Q1 | Q2 | Q3 |
| | 10 | 14 | 17 | 11 | 14 | 17 | 9 | 13 | 16 |